# Probing Electronic Correlations in Actinide Materials Using Multipolar Transitions

J.A. Bradley[1,2], S. Sen Gupta[3], G.T. Seidler[1,*], K.T. Moore[4], M.W. Haverkort[5], G.A. Sawatzky[3], S.D. Conradson[2], D.L. Clark[2], S.A. Kozimor[2], K.S. Boland[2]

*1. Physics Department, University of Washington, Seattle, WA 98195*

*2. Los Alamos National Laboratory, Los Alamos, NM, 87545*

*3. Department of Physics and Astronomy, University of British Columbia, Vancouver, BC V6T 1Z1, Canada*

*4. Lawrence Livermore National Laboratory, Livermore, CA 94550*

*5. Max Planck Institute for Solid State Research, Stuttgart, Germany, D-70569*

We report nonresonant inelastic x-ray scattering from the semi-core $5d$ levels of several actinide compounds. Dipole-forbidden, high-multipole features form a rich bound-state spectrum dependent on valence electron configuration and spin-orbit and Coulomb interactions. Cross-material comparisons, together with the anomalously high Coulomb screening required for agreement between atomic multiplet theory and experiment, demonstrate sensitivity to the neighboring electronic environment, such as is needed to address long-standing questions of electronic localization and bonding in $5f$ compounds.

The valence electronic structure of actinide (5*f*) compounds exhibits a profound competition between localizing and delocalizing influences. On-site Coulomb, exchange and spin-orbit interactions result in Hund's rules and atomic-multiplet structure, but these are matched by nontrivial crystal fields and interatomic hybridization, enabling more band-like behavior. This interplay results in strongly correlated electronic behavior and complex phase diagrams for the light actinides while also fueling considerable theoretical challenges in the field.[1-2] Consequently, while the understanding of actinide materials has grown appreciably in the last two decades, a fundamental grasp of their physical properties remains elusive.[3-4]

From the perspective of application, difficulties in both theory and experiment are clear. First, the *ab initio* design of actinide-lanthanide separation agents is key to achieving sustainable nuclear power with decreased environmental impact. Recent studies of actinide chemical bonding are leading a revision of previous notions of valence state and electronic orbital mixing, thus putting a premium on experimental methods sensitive to the valence electronic structure and low-energy electronic excited states.[5-6] Second, the highest demand is on experimental methods which are both sensitive to the local electronic environment and also compatible with model environments mimicking the extreme conditions present in reactor vessels or fuel pre- or post-processing.

Core-level spectroscopies, such as x-ray absorption spectroscopy (XAS) and electron energy loss spectroscopy (EELS), have led to successful descriptions of the electronic ground states of transition metal and rare earth materials.[7] For actinides, on the other hand, the situation is complicated by the short core-hole lifetime of tightly bound initial states, the constraints imposed by the dipole selection rule, and the practical difficulties imposed by the extreme surface sensitivity of XAS for low-energy edges.[3]

Nonresonant inelastic x-ray scattering (NIXS) satisfies the driving motivations, above, for a new experimental technique for actinide science. For low-energy edges, NIXS uses hard x-rays, so it is bulk-sensitive and compatible with extreme sample environments.[8] Furthermore, NIXS is a flexible probe of local electronic structure: Its selection rules can be tuned by variation of the momentum transfer of the scattering event.[9-10] The NIXS signal is proportional to the dynamic structure factor,

$$S(\mathbf{q},\omega) = \left|\langle f | e^{i\mathbf{q}\cdot\mathbf{r}} | i \rangle\right|^2 \delta(\Omega-\omega) = \left|\sum_{l,m} C_{lm} \langle f | Y_{lm}(\hat{\mathbf{r}}) | i \rangle\right|^2 \delta(\Omega-\omega), \quad (1)$$

where $q$ is momentum transfer, $|i\rangle$ and $\langle f|$ are the initial and final electronic states in the target, $\omega$ is photon energy loss, $\Omega$ is the energy gain of the electronic state, and there are implicit sums over individual electronic coordinates, **r**, and final states, $\langle f|$. At low momentum transfer, NIXS is a bulk-sensitive alternative to XAS, *i.e.*, both are sensitive to dipole transitions.[8] At increasing $q$, high-order angular matrix elements begin to dominate Eqn. 1, and NIXS probes final states inaccessible to dipole spectroscopy,[9-13] including states of high final angular momentum.[12] In either case, the final states probed are multi-electronic, correlated excitations. Here, we show that NIXS provides a more direct and complete characterization of the $O_{4,5}$ ($5d \rightarrow 5f$) transition region, giving new information germane to the central questions of localization, valence and excited-state electronic structure in actinide compounds.

NIXS measurements were made using the lower energy resolution inelastic x-ray scattering spectrometer at the PNC/XOR 20-ID beamline of the Advanced Photon

Source.[16] Experimental methods and data reduction follow previous methods; [17-18] in particular, incident photon energy was ~10 keV and hence measurements were purely bulk-sensitive. EELS experiments on the $O_{4,5}$ edge were performed as previously described.[14] The preparation and handling of compounds was carried out with variations on the previous reports and is described in detail elsewhere.[6] Both $UO_2$ and $U_3O_7$ were characterized by U $L_3$-edge EXAFS, x-ray and neutron powder diffraction, and O K-edge XAS—all were found to be in agreement with prior work.

In Fig. 1, we show the low-$q$ NIXS $O_{4,5}$ energy-loss spectra for $ThO_2$ and $UO_2$, along with dipole-limit EELS of both the oxides and the elemental metals. The NIXS low-$q$ spectra agree well with the EELS results. Minor differences in the $ThO_2$ spectra are due to the NIXS being slightly outside the dipole limit. Dipole transitions for the 5$d$ to 5$f$ valence states are not particularly informative because they probe largely the "giant dipole resonance" (GDR), which lacks specificity. [19-22] To wit, compare the metal to corresponding oxide spectra: They are nearly identical, though the systems are chemically quite different. Consequently, understanding of electronic structure in actinides has chiefly come from weak, dipole-forbidden features in the pre-edge for the $5d \rightarrow 5f$ transition or from other excitations, such as the $N_{4,5}$ ($4d \rightarrow 5f$) and the $M_{4,5}$ ($3d \rightarrow 5f$).[14-15, 21, 23-25]

With this in mind, in Fig. 2 we show q-dependent NIXS measurements of two $5f^0$ systems, $ThO_2$ and $Cs_2UO_2Cl_4$. The individual multipole components for the $Th^{4+}$ ion ($5f^0$) as predicted by atomic multiplet theory are plotted in Fig. 2b, and the smooth curves in Fig. 2a show the corresponding calculated $S(q,\omega)$, using the XTLS8.3 code.[26] Only odd multipole transitions ($Y_{lm}$ where $l = 1, 3, 5$) are allowed by parity for $d \to f$ transitions. These results suggest that the light actinides generally show a strong multiplet splitting of the 5d-5f excitonic states, sufficient to move the lowest energy (and, via Hund's rule, highest angular momentum) states below the continuum, and consequently into the regime of localization. The clearest evidence supporting this conclusion is the strong agreement between the atomic-based theory and experiment, especially at higher $q$. However, this accord requires unexpectedly high screening of the atomic Coulomb interaction. Coulomb matrix elements were set to only 60% of their atomic Hartree-Fock value, as opposed to the usual 80% for transition metal and rare earth systems. This suggests a broader range of theoretical approaches may be valuable in interpreting these spectra. In the atomic picture, one possible cause for the exaggerated screening is atomic configuration mixing, as investigated by Sen Gupta, *et al*. [22]

We can use the atomic picture to explain the general structure in the various high-$q$ NIXS spectra. By varying the strength of the spin-orbit coupling and the Coulomb interaction in the $Th^{4+}$ multiplet calculation, we find that the large scale, two-feature structure (~87 and 94eV for $ThO_2$, ~97 and 104eV for $Cs_2UO_2Cl_4$) of the NIXS spectra reflects 5d spin-orbit splitting, while the smaller splitting within each feature depends on the 5d-5f Coulomb interaction. Furthermore, the size of this measured spin-orbit gap is consistent with 5d electron binding energies measured by XPS for these systems.[15] This is remarkable because in dipole-limit spectroscopies, the 5d spin-orbit splitting is invisible at the $O_{4,5}$ edge due to the high intrinsic broadening of the GDR.[14]

A closer look at the spectra is required to identify the embedded atomic information. In particular, the $Th^{4+}$ calculation (Fig. 2b) predicts a mid-$q$ ($l = 3$) resonance followed by the rise of a slightly lower energy feature at highest $q$ ($l = 5$). This is observed quantitatively in both the low (~87eV) and high (~94eV) energy features in the $ThO_2$ spectra. However, it is also observed, qualitatively, in both the features that comprise the $Cs_2UO_2Cl_4$ spectrum even with the material's higher oxidation state, differing metal-ligand coordination and more covalent bonding.[5] For $ThO_2$, there is a smaller separation between the $l = 3$ and $l = 5$ transition intensities for the high-energy feature. This difference in spacing makes the high-energy feature appear to "move" with $q$. The high-energy feature in the $Cs_2UO_2Cl_4$ is more widely split, so two distinct peaks are visible at moderate $q$, giving way to a well-defined shoulder as $q$ increases.

These arguments are strengthened by examining systems with nonzero 5$f$ occupancy. In Fig. 3, we present measurements and calculations for $UO_2$ (homogeneously 5$f^2$) and $U_3O_7$ (mixed-valent). Pre-threshold excitations emerge in these spectra (at ~97 and 105eV) in a generally similar fashion to the 5$f^0$ case. However, in contrast to the 5$f^0$ case, the 5$f^2$ predictions put the $l = 3$ and $l = 5$ multipole transition intensities at the same energy for the higher energy feature (~104eV), meaning that the NIXS high energy peak position should be stable in energy with changes in $q$. This prediction is observed, meaning that feature location and motion as a function of $q$ fingerprint atomic-level properties of the system.

We now turn to the relationship between local chemistry and the NIXS $5d \rightarrow 5f$ spectra. As with the 5$f^0$ materials, anomalously high Coulomb screening (50%) is required for best agreement in the 5$f^2$ compounds, suggesting that this is a generic issue for light actinide materials. Such deviations from the isolated-atomic picture require a significant influence of the local electronic environment. Furthermore, the pre-threshold region for $U_3O_7$ differs from that of $UO_2$ by a shift (~0.5 eV higher) and broadening of the higher energy feature. First, given the +2/3 change in idealized U oxidation state, this reflects metal *f*-electronic structure that is not grossly perturbed by the extra oxygen inclusions. This supports actinide covalent bonds being more strongly 6*d* than 5*f* in character, a topic of continuing debate since the 1950's.[6] Second, the modest spectral changes that are observed are reminiscent of a recent NIXS study [12] where 4*d*-core to valence excitations in 4*f* compounds demonstrated sensitivity to local environment and bonding characteristics. The direct interrogation of 5*f* states is a key advantage of studying the $O_{4,5}$ higher multipole transitions, and suggests several future directions: Pressure dependent NIXS $O_{4,5}$ studies of heavier actinides (Am or Cm) would be a valuable and novel probe of the Mott transition,[3] while solution-phase studies would give important insight into the fundamental chemistry of the actinide/lanthanide separations process.[27]

In summary, we report new measurements and calculations for NIXS from 5*d* semi-core states to 5*f* valence states. We find that high-order multipole transitions access states localized in nature and high in orbital angular momentum. Agreement with theory is good, subject to one subtle, but important *caveat*: the necessary use of unexpectedly large screening effects. This indicates an obvious direction for future theoretical work, in that one must include the effects of local environment in a fundamental way.[22] Our results, together with the inherent compatibility of NIXS with extreme environments, strongly endorse the use of NIXS $5d \rightarrow 5f$ measurements in ongoing debates about the local electronic structure of actinide materials.

**Acknowledgments**: This work was supported by the U.S. Department of Energy and the Natural Sciences and Engineering Research Council (NSERC) of Canada Additional support came from a Frederick Reines Postdoctoral Fellowship (SAK) and a Seaborg Institute Graduate Fellowship (JAB). Measurements at the Advanced Photon Source (APS) are supported by the U.S. Department of Energy, NSERC of Canada, the University of Washington, Simon Fraser University and the APS.

**Figure Captions**

**Figure 1**: Actinide $O_{4,5}$ ($5d$) dipole- or near dipole-channel spectra derived from NIXS and EELS. [**14**] Curves have been offset vertically for clarity. Incident photon energy in the NIXS measurements is ~10 keV.

**Figure 2**: (a) Pre-threshold NIXS $O_{4,5}$ ($5d$) spectra from $ThO_2$. Points are experimental measurement, solid curves are theoretical calculations. Momentum transfers ($q$) are 3.1, 5.3, 7.7, 8.9 and 10.0 Å$^{-1}$. (b) Calculated transition intensities are resolved into multipolar components to clarify the $q$-dependence observed in NIXS. (c) NIXS $O_{4,5}$ from $Cs_2UO_2Cl_4$, similar to (a), but without $q$=3.1 Å$^{-1}$. Note that the energy scales for the two materials are different, and that an empirical energy scaling has been used to qualitatively match important features in the spectra. Curves have been offset vertically for clarity.

**Figure 3**: (a) Pre-threshold NIXS $O_{4,5}$ ($5d$) spectra from $UO_2$. Points are experimental measurement, solid curves are theoretical calculations. Momentum transfers ($q$) are 3.1, 5.3, 7.7, 8.9 and 10.0 Å$^{-1}$. (b) Calculated transition intensities for the $U^{6+}$ ion are resolved into multipolar components. (c) Pre-threshold NIXS $O_{4,5}$ spectra from $U_3O_7$ for the same $q$ as in part (a). Curves have been offset vertically for clarity.

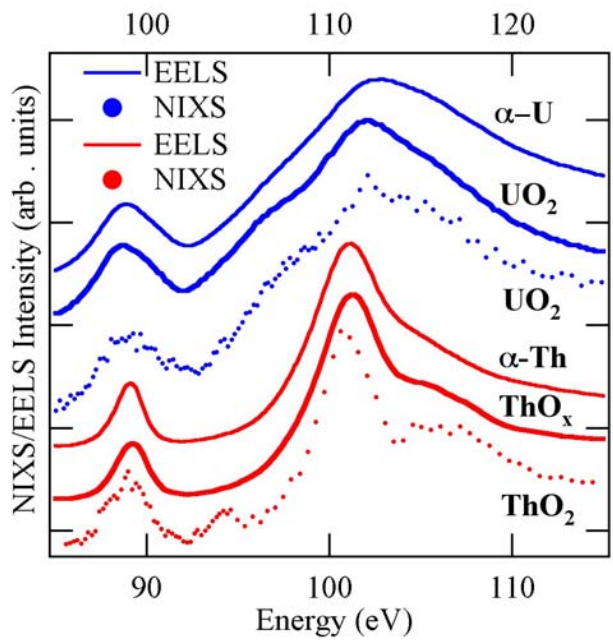

**Figure 1**: J.A. Bradley, *et al*., 2010.

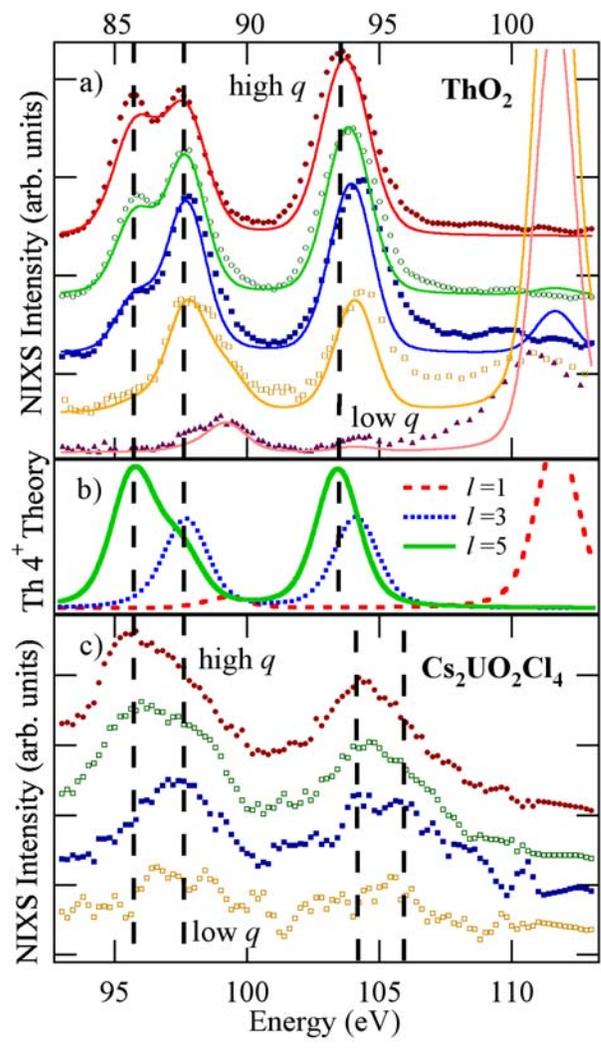

**Figure 2**: J. A. Bradley, *et al*. 2010

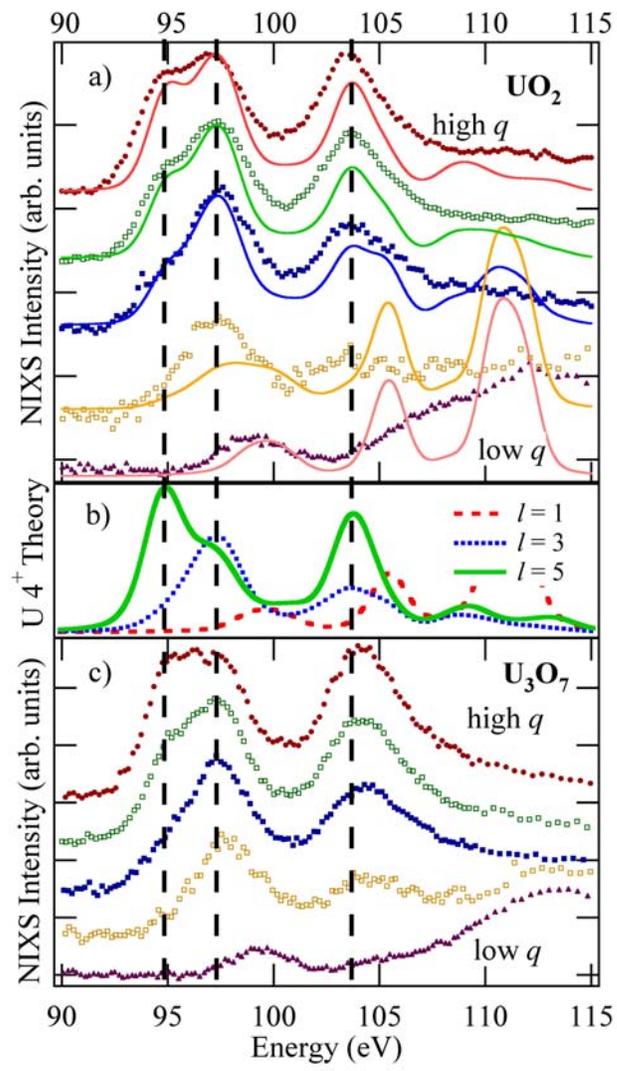

**Figure 3**: J.A. Bradley, *et al.*, 2010.